\documentclass[twocolumn]{aastex701}
\usepackage{xspace}
\usepackage{xcolor}

\defcitealias{Abbot2023}{Abbot et al.}
\defcitealias{Abbot2021}{Abbot et al.}
\defcitealias{Zeebe2015}{Zeebe}

\begin{document}

\title{Democratic heliocentric coordinates underestimate the rate of instabilities in long-term integrations of the Solar System}

\author[orcid=0000-0003-1927-731X]{Hanno Rein}
\email{hanno.rein@utoronto.ca}

\affiliation{Dept. of Physical and Environmental Sciences, University of Toronto at Scarborough, Toronto, Ontario, M1C 1A4, Canada}
\affiliation{Dept. of Physics, University of Toronto, Toronto, Ontario, M5S 3H4, Canada}
\affiliation{Dept. of Astronomy and Astrophysics, University of Toronto, Toronto, Ontario, M5S 3H4, Canada}
\author[orcid=0009-0004-8126-3642]{Kavi Dey}
\affiliation{Dept. of Physics, Harvey Mudd College, 301 Platt Blvd., Claremont 91711, USA}
\email{kdey@g.hmc.edu}
\author[orcid=0000-0002-9908-8705]{Daniel Tamayo}
\affiliation{Dept. of Physics, Harvey Mudd College, 301 Platt Blvd., Claremont 91711, USA}
\email{dtamayo@g.hmc.edu}

\begin{abstract}
    Wisdom-Holman (WH) integrators are symplectic operator-splitting methods widely used for long-term N-body simulations of planetary systems.
    Most implementations use either Jacobi coordinates or democratic heliocentric coordinates (DHC) for the Hamiltonian splitting, resulting in slightly different algorithms.
    In this paper we report results from numerical experiments, which show that integrations of the Solar System using DHC coordinates with typical timesteps of a few days suppress instabilities of the planet Mercury.
    We further show that this is due to an eccentricity dependent artificial numerical precession introduced by the DHC splitting. 
    While the DHC splitting converges to the correct results at shorter timesteps of $\sim 0.6$~days, we argue that Jacobi coordinates remain reliable to significantly longer timesteps when orbits become moderately eccentric, and are thus a better choice when the innermost planet can reach high eccentricities.
\end{abstract}


\section{Introduction} 
Numerical integrations of the Solar System, have a long history (see e.g. \citealt{Eckert1951} for one of the first simulations of this kind, and \citealt{Laskar2013} for a historical summary of the field).
Although different authors use different numerical algorithms, most modern long-term simulations use symplectic integrators that make use of the Wisdom-Holman splitting \citep{Wisdom1981, WisdomHolman1991, Kinoshita1991}.
Various variations of the Wisdom-Holman~(WH) algorithm exist, differing in the order of the scheme, the coordinate system used for the Hamiltonian splitting \citep{HernandezDehnen2016}, and other details such as whether or not symplectic correctors are used \citep[for a comparisons of methods see][]{ReinTamayoBrown2019}.

Several authors have argued that it is important for the adopted timestep to be short enough to resolve pericenter passages, which becomes challenging at high orbital eccentricities like those achieved when Mercury goes unstable \citep{Rauch1999, Wisdom2015, Hernandez2022}.

This paper shows that this criterion depends on the coordinate system used.
To clarify, we do not refer to the coordinate system in which the user inputs and outputs their positions and velocities (e.g. ecliptic vs equatorial).
Our focus is rather on the choice of coordinates used to split off a dominant Keplerian term from the full Hamiltonian as part of the WH algorithm.
The two most commonly used coordinate systems for the Hamiltonian splitting are Jacobi coordinates and democratic heliocentric coordinates (DHC).
Both coordinate systems have advantages and disadvantages.
For example, DHC \citep{Duncan1998} can be used for hybrid symplectic integrators and can be more easily parallelized \citep{Javaheri+2023}.
Jacobi coordinates on the other hand are more accurate and can be used with symplectic correctors to achieve very high order \citep{Wisdom1996, Blanes2013}.
For a precise definition and a more detailed comparison of these coordinate systems see \cite{ReinTamayo2019}.

Both coordinate systems have been used to study the long-term evolution and stability of the Solar System.
The first direct integrations over 5~Gyr were performed by \cite{LaskarGastineau2009} with the SABA integrator, a high order integrator using Jacobi coordinates \citep{LaskarRobutel2001}.
More recently, \cite{Kaib2025} used the MERCURY hybrid integrator \citep{Chambers1999} with DHC to study the influence of passing field stars on the Solar System.
Notably, the simulations that constitute the control group of \cite{Kaib2025} and should be comparable to those of  \cite{LaskarGastineau2009} show an instability rate of only 0.1\%, whereas the simulations of \cite{LaskarGastineau2009} and others \citep{Zeebe2015, Abbot2021, Abbot2023} result in an instability rate an order of magnitude larger, $\sim1$\%.

Similar discrepancies were also observed by \cite{Zeebe2015} who studied the position error of Mercury when using different Hamiltonian splittings.
\cite{Zeebe2015} correctly concluded that DHC coordinates lack accuracy and Jacobi coordinates are better suited to the problem. 

A comparison between Jacobi coordinates and heliocentric coordinates (HC) was also performed by \cite{Farres2013}.
The authors use the energy error as a metric and find that, for the same timestep, Jacobi coordinates perform one order of magnitude better in simulations of the full Solar System (see their Figure 7).
The focus of \cite{Farres2013} is on finding the most efficient method for very high accuracy simulations where the discretizations errors are of the order of machine precision.
Our goal is different. 
We want to understand if and why the stability of the Solar System depends on the coordinate system.
We argue that the error in Mercury's precession rate is better suited than the energy error to assess and interpret the reliability of long-term simulations of the Solar System).

We begin in Sec.~\ref{sec:5Gyr} by running numerical experiments which show that this discrepancy is indeed due to the different coordinate systems used in the Hamiltonian splitting.
We show that simulations using DHC require a significantly smaller timestep in order to be statistically converged.
In Sec.~\ref{sec:precession}, we demonstrate that the issue suppressing dynamical instabilities is best understood in terms of errors not in the position or energy but rather in the precession rate, as the Solar System's evolution is primarily governed by secular dynamics. 
With this understanding, we show that integrations using DHC become unreliable when a planet's eccentricity becomes large, or more precisely when the pericenter crossing timescale becomes comparable to the adopted timestep.
We show in Sec.~\ref{sec:highe} that the stabilizing effect of the artificial numerical precession is the same as the stabilizing effect due to general relativistic (GR) precession.
By contrast, the splitting using Jacobi coordinates solves the two-body problem for the inner-most planet exactly and reliably reproduces the correct precession rate out to much larger eccentricities. 
This allows for a significantly longer timestep when using Jacobi coordinates as compared to DHC.
In Sec.~\ref{sec:discussion} we discuss how the precession error originates from the integrator's jump step.
We conclude in Sec.~\ref{sec:conclusions} with some general advice.

\section{Tests} \label{sec:tests}

All simulations were performed with the REBOUND package \citep{ReinLiu2012}.
REBOUND includes several integrators, in particular the WH integrator WHFast \citep{ReinTamayo2015} which allows us to choose the coordinate system for the Hamiltonian splitting. 
The computational costs of WHFast integrations are almost the same for both Jacobi coordinates and DHC (the difference is $\sim8\%$).

We also run simulations with the WHFast512 integrator \citep{Javaheri+2023}, which is significantly faster (4.7x) but currently only supports DHC\footnote{A version of WHFast512 supporting Jacobi coordinates is currently in development. This can be found on the public branch \texttt{whfast512\_asm}.}.
We include comparisons with WHFast512 because it does not reuse any code from WHFast, and can therefore be seen as an independent implementation of a WH integrator with DHC.
This helps rule out any errors in the implementation as a cause for the observed discrepancies.

For all simulations we use the present-day positions and velocities of all eight Solar System planets as initial conditions. 
We combine each planet's mass with that of any satellites into a single point particle located at the planet-satellite barycenter.
We ignore stellar mass-loss, the solar quadrupole, tides, and all minor planets and asteroids.
We do however include general relativistic corrections in the form of an additional potential \citep{Nobili1986, Tamayo2020} as the additional precession rate is very important to correctly capture the long-term stability of the Solar System \citep{Laskar2008, LaskarGastineau2009, BoueLaskarFarago2012, BrownRein2023}.
Numerical experiments by \cite{Abbot2023} suggest that this simplified physical setup incorporates all the relevant physics to capture the instability rate of Mercury.

We determine if a system is unstable by monitoring the eccentricity of planets. 
Specifically, we define a system as unstable if an eccentricity (most likely that of Mercury) reaches 0.7 at any time during the integration.
We ran additional tests with different instability criteria, but they do not change the results significantly \cite[see also Table~1 of the Supplementary Information of][]{LaskarGastineau2009}.

\subsection{Instability rate in 5 Gyr integrations}\label{sec:5Gyr}

\begin{table*}
    \caption{Comparison of instability rates for 5~Gyr simulations of the Solar System using Jacobi (top) vs. democratic heliocentric (bottom) coordinates, using a range of integrators. 
    Integrations are sorted in descending order by timestep.
    Also shown are the generalized order and the cost (the number of interaction steps per timestep) for each integrator.
    Additional columns list
    whether the timestep was decreased once Mercury's eccentricity rose above a particular threshold (details vary by paper), as well as the number of runs in each simulation suite. 
    The final column lists the percentage of cases in which Mercury goes unstable, with errors denoting $1\sigma$ counting statistics. Integrations in Jacobi coordinates agree statistically on the instability rate over a broad range of timesteps, while the DHC integrations only converge on the correct instability rate for timesteps $\lesssim 0.6$~days. \label{tab1}}
    \centering
    \begin{tabular}{lllllcccc}
        \hline
        \hline
        Splitting & Integrator & Generalized order & Cost & Timestep & Varied & Runs & Reference & Instabil. rate\\ 
        \hline
        Jacobi &SABA    & $\epsilon dt^6 {+} \epsilon^2 dt^2$                                & 4  & 9.3 days    & Yes   & 2501  & \citeauthor*{LaskarGastineau2009} \citeyear{LaskarGastineau2009} & $0.9\%\pm0.1\%$ \\
        Jacobi &WHCKL   & $\epsilon dt^{18}{+} \epsilon^2 dt^4 {+} \epsilon^4 dt^3$          & 2  & 8.1 days    & No    & 1008   & \citetalias{Abbot2021} \citeyear{Abbot2021} & $1.2\% \pm 0.3\%$ \\
        Jacobi &WHFast  & $\epsilon dt^{2}$                                                  & 1  & 6 days      & No    & 640    & This Paper & $0.5\% \pm 0.3\%$ \\
        Jacobi &HNBody  & $\epsilon dt^{2}$                                                  & 1  & 4 days      & Yes   & 1600   & \citetalias{Zeebe2015} \citeyear{Zeebe2015} & $0.6\% \pm 0.2\%$ \\
        Jacobi &WHFast  & $\epsilon dt^{2}$                                                  & 1  & 3.16 days   & No    & 2750   & \citetalias{Abbot2023} \citeyear{Abbot2023} & $0.8\% \pm 0.2\%$ \\
        Jacobi &WHFast  & $\epsilon dt^{2}$                                                  & 1  & 3.16 days   & Yes   & 2750   & \citetalias{Abbot2023} \citeyear{Abbot2023} & $1.0\% \pm 0.2\%$ \\
        \hline                                                                                                          
        DHC &WHFast512  & $\epsilon dt^{2}$                                                  & 1  & 6.5 days    & No    & 1000   & This Paper & 0\% \\
        DHC &WHFast512  & $\epsilon dt^{2}$                                                  & 1  & 6 days      & No    & 400    & This Paper & 0\% \\
        DHC &WHFast     & $\epsilon dt^{2}$                                                  & 1  & 6 days      & No    & 960    & This Paper & 0\% \\
        DHC &MERCURY    & $\epsilon dt^{2}$                                                  & 1  & 1.5 days    & No    & 1000   & \citeauthor*{Kaib2025} \citeyear{Kaib2025} & $0.1\% \pm 0.1\%$\\
        DHC &WHFast512  & $\epsilon dt^{2}$                                                  & 1  & 0.6 days    & No    & 480    & This Paper & $1.3\% \pm 0.5\%$ \\ 
        \hline
    \end{tabular}
\end{table*}

As a first test, we run sets of 5~Gyr integrations of the Solar System with different integrators, Hamiltonian splittings, and timesteps and measure the instability rate in each set.
Each simulation is slightly perturbed by changing the position of Earth on the order of a few meters to obtain a set of equally valid realizations.
The results are shown in Table~\ref{tab1}, together with previous results in the literature. 

We see that integrations using Jacobi coordinates all agree with the instability rate of $\sim 1\%$ determined by \cite{LaskarGastineau2009}.
However, our simulations with DHC and timesteps of $\sim6$~days suppress all instabilities.
The fact that we reproduce this result with different integrators and agree with the integrations of \cite{Kaib2025} is reassuring as it shows this effect is not due to a specific implementation.

Importantly, we are able to reproduce the correct instability rate with DHC simulations that use an order of magnitude smaller timestep of $0.6$~days (bottom row).
This suggests that the issue at hand is a simple convergence issue that can be resolved by reducing the timestep.
We will show more evidence for this interpretation in the next section.

\subsection{Artificial numerical precession}\label{sec:precession}

\begin{figure*}[t]
\includegraphics[width=\textwidth]{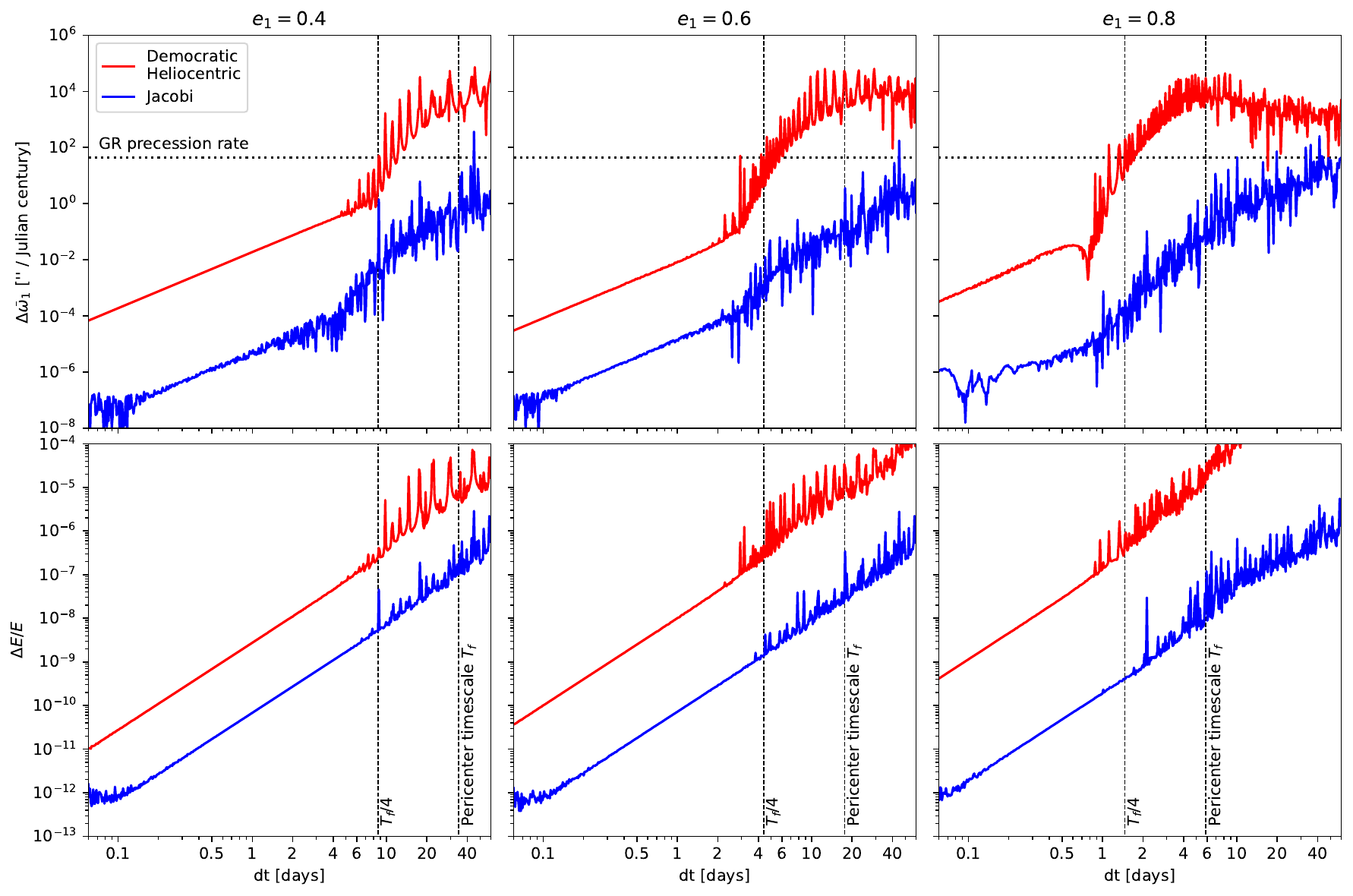} 
\caption{Top: Error in Mercury's precession rate as a function of timestep for different eccentricities of Mercury.
    The blue curve shows the error in simulations using Jacobi coordinates. The red curve shows the error in simulations using democratic heliocentric coordinates. 
    Also shown are the pericenter timescale of Mercury $T_f$ as well as $T_f/4$ as vertical lines and the precession rate of Mercury due to general relativity as a horizontal line. 
    Bottom: Same as in the top row but showing the relative energy error.
    \label{fig:prec}}
\end{figure*}

We next run sets of simulations for $10^3$~yrs with both DHC and Jacobi coordinates. 
We again use the present day values for all initial orbital parameters except for Mercury's eccentricity $e_1$ which we adjust manually this time. 
To measure the errors of different schemes, we compare them against integrations using the high-order, adaptive-timestep IAS15 integrator \citep{ReinSpiegel2015}, which obtains correct results to within accumulated floating-point errors.

In Fig.~\ref{fig:prec} we plot the magnitude of the error in the precession rate of Mercury (top) and the magnitude of the relative energy error (bottom) as a function of the timestep for the three cases of $e_1=0.4$, $0.6$, and $0.8$.
We also plot the precession rate of Mercury due to general relativistic effects (horizontal line) and the timescale for Mercury to cross the pericenter (vertical line) defined as:
\begin{eqnarray}
    T_f &\equiv& \frac{2\pi}{n_1} \frac{(1-e_1)^2}{\sqrt{1-e_1^2}},    \label{eq:peri}
\end{eqnarray}
where $n_1$ is the mean motion of Mercury \citep{Wisdom2015}.
Note that the pericenter timescale can be significantly shorter than the orbital period for eccentric orbits.

For small timesteps, the numerical discretization error (or method error) of the WH integrator is proportional to the timestep squared as expected for a second order scheme.
One can see that for very small timesteps, $dt<0.1$~days, the integrator using Jacobi coordinates reaches a floor which is dominated by round-off errors due to finite floating point precision.
For large eccentricities and timesteps, the errors in the precession rate as well as the energy errors show signs of timestep resonances~(see Sect.~\ref{sec:discussion}).

One can see that the WH integrator with Jacobi coordinates maintains a precession error that is small compared to the precession resulting from GR for timesteps $\lesssim 10$~days. 
On the other hand, the WH integrator with DHC introduces an artificial numerical precession which is strongly dependent on both the timestep and the eccentricity of Mercury. 
The error in the preccesion rate can be significantly larger than the GR precession rate, even for timesteps as small as 1~day.

Previous studies have shown that the instability fraction is sensitive to GR and any additional precession \citep{Laskar2008, LaskarGastineau2009, BoueLaskarFarago2012, BrownRein2023}.
We therefore should expect a significant effect on the stability rate when the error in the precession rate become comparable to the GR precession rate. 
We note that the artificial numerical precession has the same direction as the physical GR precession, and that the breakdown of the WH integrator with DHC occurs when the timestep becomes comparable to the pericenter-crossing timescale, $dt \gtrsim T_f/4$.  
We thus confirm our earlier claim that the discrepancy observed in long-term integrations is due to the non-convergence of simulations with DHC.
Note that using the energy error as the lone metric for the accuracy of these secularly evolving simulations would not have allowed us to come up with this physical interpretation.

\subsection{A physical cause for the different instability rates}\label{sec:highe}

\begin{figure*}
\includegraphics[width=\textwidth]{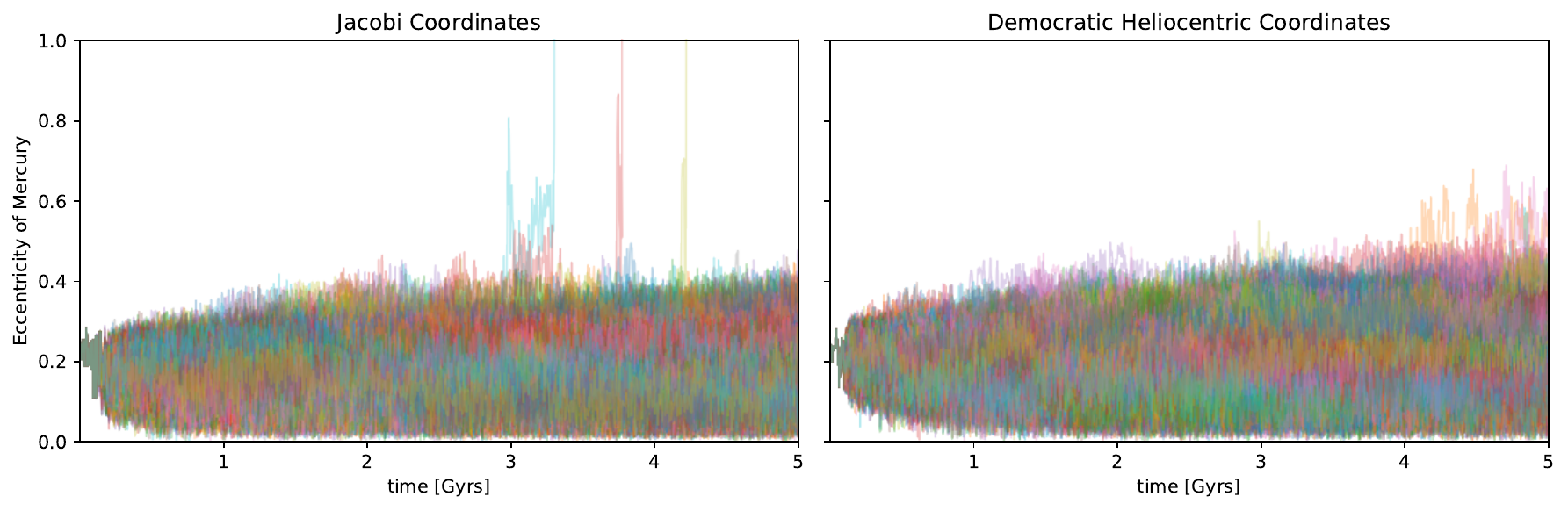} 
\caption{Left panel shows Mercury's eccentricity as a function of time in 640 integrations using Jacobi coordinates, exhibiting near vertical take-offs when the frequency dominantly associated with Mercury's eccentricity precession $g_1$ becomes resonant with that of Jupiter $g_5$. Right panel shows the same for 640 integrations using democratic heliocentric coordinates, where the vertical takeoffs are suppressed by artificial precession (see text). Both sets of integrations use a 6 day timestep. \label{fig:e}}
\end{figure*}

Our results above suggest that simulations using DHC only break down when eccentricities reach large values. 
To explore this further, we take one of the unstable Solar System integrations and restart it 5~Myrs prior to the eccentricity of Mercury reaching large (0.7) values.
We then integrate it forward in time for 20~Myrs.
We once again run simulations with different coordinate systems and timesteps.
This time we also vary the general relativistic precession rate artificially from 0 (no GR precession), to 1 (normal GR precession), to 2 (double the GR precession). 
We run 100 simulations for each case and slightly perturb again the position of Earth to obtain a set of equally valid realizations.

This setup lets us quickly measure the instability rate for an ensemble of simulations.
We chose to restart simulations 5~Myrs before the instability, roughly one Lyapunov timescale \citep{Laskar1989}, to give individual simulations time to diverge slightly through phase space.
If we were to restart them much earlier, then the small perturbations we add to the initial conditions would move most trajectories away from the high eccentricity phase we aim to explore.

\begin{table}
\caption{Instability rate in simulations restarted just prior to a phase of high eccentricity in Mercury's orbit.\label{tab2}}
\centering
\hskip-1.4cm\begin{tabular}{lccc}
Coordinates & Timestep & GR & Instability rate\\
\hline \hline
Jacobi & 6 days & 1.0 & 100\% \\
Jacobi & 6 days & 2.0 & 0\% \\
Jacobi & 6 days & 0.0 & 100\% \\
DHC & 6 days & 1.0 & 0\% \\
DHC & 0.6 days & 1.0 & 100\% \\
\hline
\end{tabular}
\end{table}

Our results are shown in Table~\ref{tab2}.
First, note that just as in Sect.~\ref{sec:5Gyr}, when using a 6~day timestep, DHC integrations again suppress all instabilities.
The instability rate of Jacobi coordinates is only recovered when the timestep for DHC is reduced to 0.6~days.

Second, the runs with Jacobi coordinates and double the GR precession rate yield no instabilities. 
This reaffirms that a change in the precession rate of order $\sim40$\arcsec/Julian century is sufficient to alter the instability rate \citep{LaskarGastineau2009, BoueLaskarFarago2012, BrownRein2023}.
When turning off GR precession completely in the simulations with Jacobi coordinates, we still see an instability rate of 100\%.
This shows that the sign of this change matters.
Moreover, because the artificial numerical precession present in DHC simulations is in the same direction as the GR precession, we can conclude that the artificial precession is making the simulations more stable in the same way that the GR precession does.

We can connect this directly to the dynamics behind these instabilities.
Chaos causes the precession frequencies of the secular modes in the Solar System to diffuse on Gyr-timescales, particularly the frequency $g_1$ associated with Mercury \citep[e.g.,][]{Hoang2021}.
\cite{BoueLaskarFarago2012} showed that Mercury is lost when $g_1$ diffuses from its present day value of $\sim 560$\arcsec/Julian century down to $\sim 425$\arcsec/Julian century, where the separatrix of the secular $g_1-g_5$ resonance appears, and Mercury's eccentricity approaches unity effectively instantaneously.
This rapid evolution on the secular timescale, which is much shorter than the Gyr timescale for chaotic diffusion, gives rise to the near-vertical unstable trajectories in the left panel of Fig.~\ref{fig:e} where we plot the eccentricity of Mercury as a function of time in simulations using Jacobi coordinates (left) and DHC (right). 

As $g_1$ approaches resonance in the DHC integrations and the eccentricity of Mercury starts to rise, the eccentricity-dependent artificial precession (Fig.~\ref{fig:prec}) raises $g_1$ and detunes it away from resonance, removing the separatrix and suppressing further eccentricity growth.
In the right panel of Fig.~\ref{fig:e} we see several instances where the eccentricity of Mercury starts to rise, only to fall back down as the artificial precession moves the system away from resonance.
This self-limiting behaviour thus shuts down the primary pathway to Mercury's instability.

\section{Discussion} \label{sec:discussion} 
Our results from Sec.~\ref{sec:tests} show clear evidence that Jacobi coordinates perform better than DHC in Solar System integrations.
This is because when using Jacobi coordinates for the Hamiltonian splitting, Mercury's motion is decomposed into a Keplerian step on a two-body orbit around the Sun, and an interaction step with the other planets.
In the limit where Mercury has a high eccentricity, the Kepler solver still correctly solves for the Keplerian motion around the Sun-Mercury barycenter for arbitrarily large timesteps.
And because both the orbital timescales and the timescales for planet-planet interactions are much longer than the pericenter timescale~$T_f$, no significant errors are expected by not resolving~$T_f$ (see Fig.~\ref{fig:prec}). 

DHC on the other hand split the Hamiltonian into three parts: Keplerian motion, planet-planet interactions, and a so-called jump step or solar term \citep[see][]{Duncan1998,ReinTamayo2019}.
The jump step depends on Mercury's momentum (and that of the other planets) relative to the Solar System barycenter. 
The Keplerian motion solved for in the Kepler step is also different from the Jacobi case. 
Mercury does not orbit the Mercury-Sun barycenter as in the Jacobi case, but rather the Solar System barycenter\footnote{More precisely, DHC uses heliocentric positions and barycentric velocities. See \cite{Duncan1998,ReinTamayo2019}.}.
In the limit of high eccentricity, the difference between Mercury's true trajectory and the Keplerian orbit the Kepler solver solves for becomes significant, especially near pericenter. 

The error of a symplectic integrator (or more generally an operator splitting method) can also be understood in terms of an infinite sum of commutators consisting of compositions of the integrator's individual steps \citep[see e.g.][]{SahaTremaine1992, Rein2019, Tamayo2020}.
The commutators involving the jump step lead to error terms not present in the case of Jacobi coordinates and it is those commutators which lead to the artificial numerical precession.
For small timesteps, these truncation (method) errors scale proportionally to the timestep squared because we have a second order method (the smooth parts of the curves in Fig.~\ref{fig:prec}).
For large timesteps, we see comb-like structures which are due to strong timestep resonances \citep{Rauch1999}.
To suppress these timestep resonances and keep the artificial numerical precession rate below that of GR in simulations using DHC we need to resolve $T_f$ with at least a few ($\sim4$) timesteps.

The effect we discuss primarily affects the innermost planet, i.e.~Mercury in the case of the Solar System.
Venus, for example, has a 2.6 times longer orbital period than Mercury.
We are guaranteed to resolve Venus' pericenter timescale up to eccentricities of $e_2\sim0.4$ as long as we resolve Mercury's orbital period.
Furthermore, as soon as the eccentricity of Venus reaches $e_2 \sim0.4$ it crosses orbits with Mercury at which point close encounters can occur and all of the fixed timestep integrators we discuss break down.

\section{Conclusions} \label{sec:conclusions} 
Different instability rates of the Solar System have been reported in the literature, ranging from $\sim0.1$\% to 1\% \citep{LaskarGastineau2009, Zeebe2015, Abbot2021, Abbot2023, Kaib2025}.
In this paper we traced this discrepancy back to the coordinate system used in the Hamiltonian splitting of the Wisdom-Holman integrator. 
We showed that simulations using democratic heliocentric coordinates (DHC) require a significantly smaller timestep in order to be statistically converged compared to simulations using Jacobi coordinates. 
The reason for this slow convergence is the artificial numerical precession introduced by DHC which is strongly eccentricity and timestep dependent.  

Our results show that DHC should not be used in simulations which require high accuracy during times when the eccentricity of planets is high. 
The Solar System is one example where high accuracy is required to measure the correct instability rate because the eccentricity of Mercury becomes large ($\gtrsim 0.4$) just before instabilities occur and numerical errors can suppress this behaviour.

Reducing the timestep in simulations using DHC is one possible solution. 
Another, pursued by \cite{Levison2000} and \cite{Lu2024} is to shift the problematic terms in the jump step during close pericenter passages into the Keplerian step and integrate them using, for example, a Bulirsch-Stoer integrator \citep{Chambers1999}.
However, we note that a simple scheme using Jacobi coordinates can reproduce the correct instability rate for surprisingly large timesteps \citep{Rein2025}.
As a result, statistically converged simulations with DHC are significantly more computationally expensive than simulations using Jacobi coordinates so that Jacobi coordinates are advantageous in most scenarios.

Most recent direct numerical integrations of planetary systems use either Jacobi coordinates or DHC, which is why we focus on these two splitting.  
However, other splittings of the Hamiltonian exists \citep{Farres2013,HernandezDehnen2016}.
In fact, there are an infinite number of ways to split the Hamiltonian\footnote{One can add and subtract any arbitrary term to the Hamiltonian.}. 
Worth pointing out are heliocentric coordinates (HC) originally introduced by Poincaré and used in a perturbative expansion by \cite{LaskarRobutel1995}. 
For direct N-body simulations \cite{Farres2013} find no significant differences between HC and DHC (see their Appendix~C).
Among the infinite possible choices of coordinate systems, Jacobi coordinates stand out as the only one which, by construction, respect the hierarchical nature of stable planetary systems.
DHC or HC have their uses in systems which are not hierarchical at all times, such as systems with a large number of particles in a disk or systems in which planets have close encounters.

We have focused our experiments and discussion on the Solar System. 
However, the general advice for running simulations of any planetary system where one wants to integrate until an instability occurs is the same:
Jacobi coordinates perform significantly better.
As soon as orbit crossing occurs and planets can have close encounters, Jacobi coordinates are no longer suitable.
Often at this point a simulation can be stopped and the system can be considered unstable. 
But if it is necessary to integrate further, then a switch to a hybrid symplectic integrator or a brute force non-symplectic integrator can be made at this point. 
Since hybrid integrators work in DHC, it is important to check that the timestep is small enough, at the very least $dt\lesssim T_f/4$, or ideally $dt\lesssim T_f/17$ as suggested by \cite{Wisdom2015}. 

\begin{acknowledgments}
    We thank an anonymous referee for thorough and helpful feedback.
    We thank Nathan Kaib, Sean Raymond, and Garett Brown for helpful discussions.
    We are also grateful for Muhammed Sajanlal's help with a literature review.
    This research has been supported by the Natural Sciences and Engineering Research Council (NSERC) Discovery Grant RGPIN-2020-04513.
\end{acknowledgments}


\bibliography{full}{}
\bibliographystyle{aasjournalv7}

\end{document}